\documentstyle[prl,aps,multicol,epsfig]{revtex}
\input{epsf}
\begin{document}
\draft
\title{On the stability of classical chaotic motion 
under system's perturbations}
\author{Giuliano Benenti$^{(1)}$, Giulio Casati$^{(1,2)}$, 
and Gregor Veble$^{(1,3)}$}  
%\affiliation{$^{(1)}$International Center for the Study of Dynamical 
%Systems, Universit\`a degli Studi dell'Insubria and} 
%\affiliation{Istituto Nazionale per la Fisica della Materia, 
%Unit\`a di Como, Via Valleggio 11, 22100 Como, Italy}   
%\affiliation{$^{(2)}$Istituto Nazionale di Fisica Nucleare, 
%Sezione di Milano, Via Celoria 16, 20133 Milano, Italy}   
\address{$^{(1)}$International Center for the Study of Dynamical 
Systems, Universit\`a degli Studi dell'Insubria and} 
\address{Istituto Nazionale per la Fisica della Materia, 
Unit\`a di Como, Via Valleggio 11, 22100 Como, Italy}   
\address{$^{(2)}$Istituto Nazionale di Fisica Nucleare, 
Sezione di Milano, Via Celoria 16, 20133 Milano, Italy}   
\address{$^{(3)}$Center for Applied Mathematics and Theoretical 
Physics, University of Maribor, Krekova ul. 2, SI-2000 Maribor,
Slovenia}
\date{December 5, 2002}
\maketitle

\begin{abstract} 
We study in detail the time behavior of classical 
fidelity for chaotic systems. 
We show in particular that the asymptotic decay, 
depending on system dynamical properties,  
can be either exponential, with a rate determined by 
the gap in the discretized Perron-Frobenius operator, 
or algebraic, with the same power as for correlation 
functions decay. Therefore the decay of 
fidelity is strictly connected to correlations decay. 
\end{abstract} 
%\pacs{05.45.-a, 05.45.Pq}  
%05.45.-a Nonlinear dynamics and nonlinear dynamical systems 
%05.45.Pq Numerical simulations of chaotic models

\begin{multicols}{2}
\narrowtext

As it is known, exponential separation of orbits starting 
from slightly different initial conditions has been 
associated to classical chaos.
It has been remarked that the situation in quantum mechanics 
is drastically different. Indeed the scalar product of 
two states $\langle\psi_1|\psi_2\rangle$ is time independent. 
This has led to the introduction of fidelity as a measure of 
stability of quantum motion \cite{peres}. 
More precisely one considers  
the overlap of two states which, starting from the same 
initial conditions, evolve under two slightly different 
Hamiltonians $H_0$ and $H_\epsilon = H_0 +\epsilon V$.
The fidelity is then given by
$ f(t) = |\langle\psi|\exp(iH_\epsilon t/\hbar) 
\exp(-iH_0 t/\hbar)|\psi\rangle|^2$.
The quantity $f(t)$ can be seen as a measure of the 
so-called Loschmidt echo: a state $|\psi\rangle$  
evolves for a time $t$ under the Hamiltonian $H_0$, then the 
motion is reversed and evolves back for the same time $t$ 
under the Hamiltonian $H_\epsilon$ and the overlap with the 
initial state $|\psi\rangle$ is considered.

However, we would like to stress that, in principle, 
such difference between classical and quantum mechanics 
actually does not exist. The Liouville equation, which 
describes classical evolution, is unitary and reversible as  
the Schr\"odinger equation. However, as stressed in several 
occasions (see, e.g., Ref. \cite{cambridge}), 
there exist time scales up 
to which quantum motion can share the properties of 
classical chaotic motion including local exponential 
instability. Due to the existence of such time scales, what 
may be different, and indeed it is, is 
the degree of stability of dynamical motion. Indeed, as 
clearly illustrated in the analysis of Loschmidt echo 
in \cite{arrow}, quantum motion turns out to be more 
stable than the classical motion.

The growing interest in quantum computers has attracted 
recent interest in this quantity as a measure of the 
stability of quantum computation in the presence of hardware 
imperfections or noisy gates operations. Confining ourselves 
to classically chaotic systems, the emerging picture which 
results from analytical and numerical investigations 
\cite{jalabert,pastawski,jacquod,tomsovic,prosen,felix,zurek,cohen,PRE} 
is that both exponential and Gaussian decay are present in 
the time behavior of fidelity. The strength of the perturbation 
determines which of the two regimes prevails. The decay rate 
in the exponential regime appears to be dominated either 
by the classical Lyapunov exponent or, according to 
Fermi golden rule, by the spreading width of the 
local density of states.

In addition, at least for short times, the decaying
behavior depends on the initial state (coherent state, 
mixture, etc.). 
While it can be true that, for practical 
purposes, the short time behavior of fidelity may be the 
most interesting one, it is also true, without any doubt, 
that in order to have a clear theoretical understanding 
and identify a possible universal type of quantum decay 
one needs to consider the asymptotic behavior of fidelity.

This arises the problem of understanding  the corresponding 
classical decay of fidelity and later on inquiring about 
the times scales at which quantum decay mimics 
the classical one. In the present paper we concentrate 
our attention on the classical behavior. 

What do we know about the decay of classical fidelity for 
chaotic systems? What is the relation with correlation 
functions? Can we derive the decay of fidelity 
from the behavior of correlations or is fidelity 
a completely independent function? 
In a recent paper \cite{PRE} it has been found 
numerically that, after an initial transient, classical 
fidelity decays exponentially and 
the rate is given by the Lyapunov exponent 
(see also \cite{prosen,eckhardt}). 
This is in also 
in agreement with previous papers 
\cite{jalabert,pastawski,jacquod} indicating that 
quantum fidelity, for strong enough quantum perturbation 
(which, for a fixed classical perturbation strength, corresponds 
to semiclassical region), decays exponentially with a rate 
given by the Lyapunov exponent of the corresponding classical 
system. 
On the other hand we know that the decay of correlation 
functions is not ruled by Lyapunov exponent. In the first 
place there is the general 
phenomenon of long time tails which means power law decay. 
In addition, for the special cases in which one can prove 
exponential decay, the rate is determined by the gap in 
the discretized Perron-Frobenius operator and not by the 
Lyapunov exponent.

In this paper, we show that the asymptotic decay of classical 
fidelity for chaotic systems is not related to the 
Lyapunov exponent: Similarly to correlation functions, 
this decay can be either exponential or power law. In 
the first case, the decay rate is determined by the gap 
in the discretized Perron-Frobenius operator, in the latter 
case the power law has the same exponent as for correlation 
functions.    

The classical fidelity $f(t)$ is defined as follows: 
\begin{equation} 
f(t)=\int_{\Omega}d{\bf x} 
\rho_\epsilon({\bf x},t)
\rho_0({\bf x},t),
\label{fido} 
\end{equation} 
where the integral is extended over the phase space, and
\begin{equation} 
\rho_0({\bf x},t)=U_0^t \rho({\bf x},0),
\quad 
\rho_\epsilon({\bf x},t)=U_\epsilon^t \rho({\bf x},0)
\end{equation} 
give the evolution after $t$ steps of the initial density 
$\rho({\bf x},0)$ 
(assumed to be normalized, i.e. 
$\int d{\bf x}\rho^2({\bf x},0)=1$) as determined 
by the $t$-th iteration of the Frobenius-Perron
operators $U_0$ and $U_\epsilon$, corresponding 
to the Hamiltonians $H_0$ and $H_\epsilon$, respectively. 
The above definition can be shown to correspond to 
the classical limit of quantum fidelity 
\cite{prosen,zurek}. 
In the ideal case of perfect echo ($\epsilon=0$), the 
fidelity does not decay, $f(t)=1$.
However, due to chaotic dynamics, when $\epsilon\ne 0$
the classical fidelity decay sets in after a time scale 
\begin{equation}
t_\nu\sim\frac{1}{\lambda}
\ln\left(\frac{\nu}{\epsilon}\right),
\label{echo}
\end{equation}
required to amplify the perturbation up to the size 
$\nu$ of the initial distribution, with $\lambda$ 
Lyapunov exponent. Thus, for $t\gg t_\nu$ the recovery of 
initial distribution via the imperfect time-reversal 
procedure fails, 
and the fidelity decay is determined 
by the decay of correlations for a 
system which evolves forward in time according to 
the Hamiltonians $H_0$ (up to time $t$) and 
$H_\epsilon$ (from time $t$ to time $2t$). 
This is conceptually similar to the ``practical'' 
irreversibility of chaotic dynamics: due to the 
exponential instability, 
any amount of numerical error in computer 
simulations rapidly effaces the memory of the 
initial distribution \cite{arrow}. 
In the present case, the coarse-graining 
which leads to irreversibility is due not to round-off 
errors but to a perturbation in the Hamiltonian.  

In the following, we illustrate this general phenomenon 
in standard models of classical chaos, characterized by 
uniform exponential instability (the sawtooth map), 
marginal stability (the stadium billard), or 
mixed phase space dynamics (the kicked rotator). 
 
The sawtooth map is defined by 
\begin{equation} 
\overline{p}=p+F_0(\theta), \quad 
\overline{\theta}=\theta +\overline{p}, 
\label{saw} 
\end{equation} 
where $(p,\theta)$ are conjugated action-angle 
variables, $F_0=K_0(\theta -\pi)$, and the 
overbars denote the variables after one map iteration. 
We consider this map on the torus 
$0\leq \theta < 2\pi$, $-\pi L\leq p <\pi L$, 
where $L$ is an integer. 
For $K_0>0$ the motion is completely chaotic and 
diffusive, with Lyapunov exponent given by 
$\lambda=\ln\{(2+K_0+[(2+K_0)^2-4]^{1/2})/2\}$.
For $K_0>1$ one can estimate the diffusion coefficient
$D$ by means of the random phase approximation, 
obtaining $D\approx (\pi^2/3)K_0^2$.  
In order to compute the fidelity (\ref{fido}), we choose  
to perturb the kicking strength $K=K_0+\epsilon$, with 
$\epsilon\ll K_0$. In practice, we follow the evolution 
of $10^8$ trajectories, which are uniformly distributed 
inside a given phase space region of area $A_0$ at time 
$t=0$. The fidelity 
$f(t)$ is given by the percentage of trajectories that 
return back to that region after $t$ iterations of the 
map (\ref{saw}) forward, followed by the backward 
evolution, now with the perturbed strength $K$, 
in the same time interval $t$. In order to study 
the approach to equilibrium for fidelity, we 
consider the quantity 
%\begin{equation} 
$g(t)=(f(t)-f(\infty))/(f(0)-f(\infty))$;
%\end{equation} 
in this way $g(t)$ drops from $1$ to $0$ when 
$t$ goes from $0$ to $\infty$. We note that 
$f(0)=1$ while, for a chaotic system, 
$f(\infty)$ is given by the ratio 
$A_0/A_c$, with $A_c$ the area of the 
chaotic component to which the initial 
distribution belongs.  

\begin{figure}
%\hspace{-0.5cm}\includegraphics[width=9cm]{fig1.eps}
\centerline{\epsfxsize=9cm\epsffile{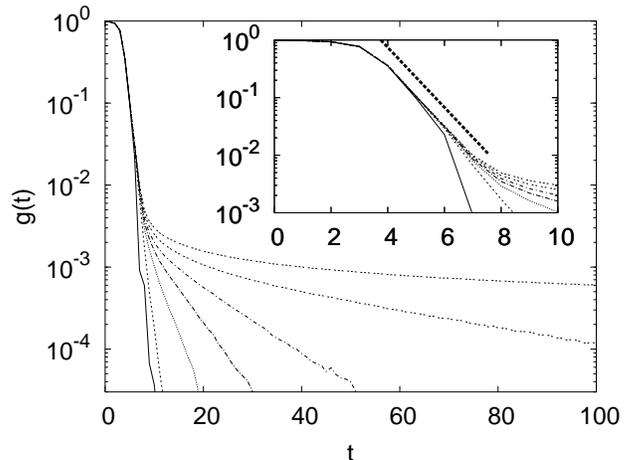}}
\caption{Decay of the fidelity $g(t)$ for the sawtooth map with 
the parameters $K_0=(\sqrt{5}+1)/2$ and $\epsilon=10^{-3}$ for 
different values of $L=1,3,5,7,10,20,\infty$ from the fastest to 
the slowest decaying curve, respectively. The initial phase space 
density is chosen as the characteristic function on 
the support given by the 
$(\theta,p)\in[0,2 \pi)\times[-\pi/100,\pi/100]$.
Note that between the Lyapunov decay and the exponential 
asymptotic decay there is a $\propto 1/\sqrt{t}$ decay, 
as expected from diffusive behavior. 
Inset: magnification of the same plot for short times, 
with the corresponding Lyapunov
decay indicated as a thick dashed line. \label{fig:fidelity}}
\end{figure}

The behavior of $g(t)$ is shown in Fig. 1, 
for $K_0=(\sqrt{5}+1)/2$ and different $L$ values. 
One can see that only the short time decay is  
determined by the Lyapunov exponent. It takes 
place for $t_\nu < t < t_\epsilon$, with 
$t_\nu$ defined in Eq.~(\ref{echo}) and 
$t_\epsilon=(1/\lambda)\ln(2\pi/\epsilon)$ time 
scale required to amplify the effect of the 
Hamiltonian perturbation up to the maximum 
extension in the angle $\theta$.
The Lyapunov regime is followed by a power law decay 
\cite{PRE} $\propto 1/\sqrt{D t}$ until the diffusion 
time $t_D\sim L^2/D$ and then the asymptotic 
relaxation to equilibrium takes place 
exponentially, with a decay rate $\gamma$  
(shown in Fig. 2), which, as discussed below,
is ruled not by the Lyapunov 
exponent but by the largest Ruelle-Pollicott 
resonance \cite{ruelle}. In particular, it is 
$\epsilon$-independent.   

\begin{figure}  
\centerline{\epsfxsize=9cm\epsffile{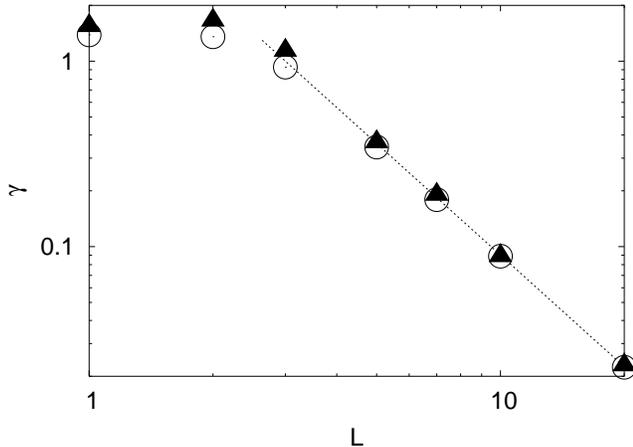}}
\caption{Asymptotic exponential decay rates of fidelity for 
the sawtooth map ($K_0=(\sqrt{5}+1)/2$, $\epsilon=10^{-3}$) as a function
of $L$. The rates are extracted by fitting the tails of the 
fidelity decay in the Fig. \ref{fig:fidelity} (triangles) and
from the discretized Perron-Frobenius operator (circles). 
The line denotes the $\propto 1/L^2$ behavior
of the decay rates, predicted by the Fokker-Planck equation,
which describes the classical motion in the diffusive regime.
\label{fig:decrates}}
\end{figure}

We determine numerically these resonances for the 
sawtooth map using the following method 
\cite{haake,fishman}: 
\newline 
(i) the phase space torus ($0\leq \theta <2\pi$, 
$-\pi L\leq p <\pi L$) is divided in $N\times NL$ square 
cells;
\newline 
(ii) the transition matrix elements between cells 
are determined numerically by iterating for one map step 
the phase space distributions given by the characteristic 
functions of each cell: in this way we build a finite 
dimensional approximation of the one-period evolution 
operator $U_0$; 
\newline 
(iii) this truncated evolution matrix $U_0^{(N)}$ 
(of size $LN^2\times LN^2$) is diagonalized: 
it is no longer unitary, and its eigenvalues 
$z_i^{(N)}$ are inside the unit circle in the 
complex plane (an example is shown in Fig. 3). 
The non-unitarity of the coarse-grained 
evolution is due to the fact that the transfer of 
probability to finer scale structures in the phase space 
is cut-off, and this results in an effective dissipation 
\cite{haake}; 
\newline 
(iv) resonances correspond to ``frozen'' non-unimodular 
eigenvalues, namely $z_i^{(N)}\to z_i$ when $N\to\infty$, 
with $|z_i|<1$. Convergence of eigenvalues 
to values inside the unit circles comes
from the asymptotic self-similarity of 
chaotic dynamics \cite{haake}. 
\newline 
As it is known, the asymptotic ($t\to\infty$) relaxation of 
correlations is determined by the resonance 
with largest modulus, $|\tilde{z}|=
{\rm max}_i|z_i|<1$, 
giving a decay rate $\gamma_0=\ln |\tilde{z}|$.  
In Fig. 2 we illustrate the good agreement between the asymptotic 
decay rate of fidelity (extracted from the data of Fig. 1) and 
the decay rate as predicted by the gap in the discretized 
Perron-Frobenius spectrum. 
It should be stressed that, since fidelity involves forward and 
backward evolution, the fidelity decay at time $t$ has to be 
compared with the correlations decay at time $2t$. For this reason 
in Fig. 2 the circles correspond to $\gamma = 2 \gamma_0$. 

\begin{figure}
\centerline{\epsfxsize=9cm\epsffile{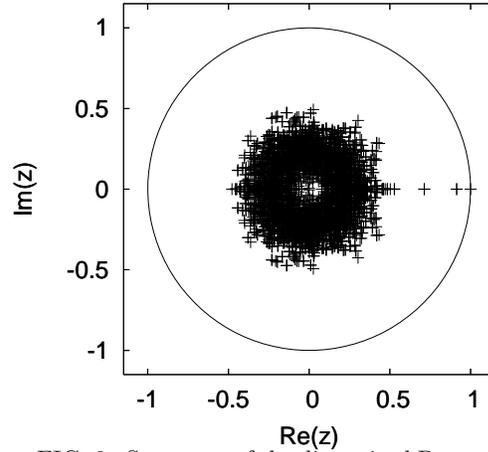}}
\caption{Spectrum of the discretized Perron-Frobenius 
operator for the sawtooth
map with parameters $K_0=(\sqrt{5}+1)/2$, $L=7$ and 
discretization $N=20$. 
The asymptotic decay of fidelity is determined by the 
largest modulus eigenvalue apart from the eigenvalue $1$.
\label{fig:perfrob}}
\end{figure}
\vspace{-0.8cm}  
\begin{figure}
\centerline{\epsfig{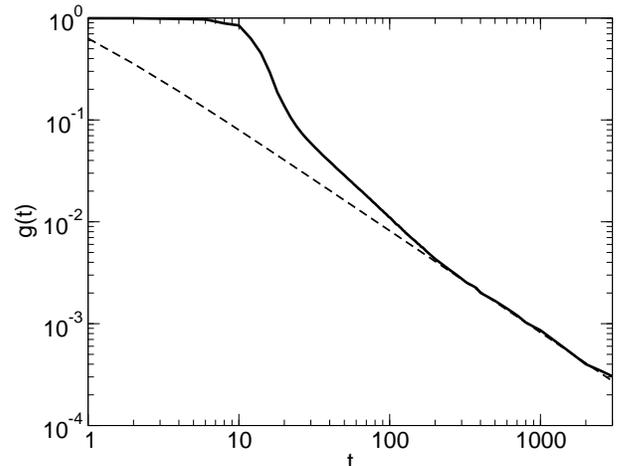}}
%\centerline{\epsfxsize=8cm\epsffile{fig4.eps}}
\caption{Power law decay of fidelity for the stadium billiard 
with radius $R=1$ and length of the straight segments 
$d_0=2$ (the perturbed stadium has $d=d_0+\epsilon$, with 
$\epsilon=2\times 10^{-3}$). The initial
phase space density was chosen to be a direct product of a 
characteristic function on a circle in configuration space, 
the center of which was at (0.5,0.25) as measured
from the center of the billiard and its radius was 0.1,
while for momenta the $\delta(|{\bf p}|-1)$ distribution was used. 
The dashed line has slope $-1$.
\label{fig:stadfid}}
\end{figure}

We would like to stress that the same qualitative 
behavior of Fig. 1 is obtained in the presence of 
stochastic noise, e.g. the backward evolution 
is driven by a kicking strength $K(t)=K_0+\epsilon(t)$, 
with $\{\epsilon(t)\}_{t=1,2,...}$ uniformly and randomly 
distributed inside the interval $[-\epsilon,\epsilon]$.
In particular, we observed the initial Lyapunov decay 
and the asymptotic exponential relaxation with the same 
rate $\gamma$. 
This means that the effect of a noisy environment 
on the decay of fidelity for a classical chaotic system 
is similar to that of a generic static Hamiltonian 
perturbation.   

\begin{figure}
%\includegraphics[width=9cm]{fig5.eps}
%\centerline{\epsfxsize=8cm\epsffile{fig5.eps}}
\centerline{\epsfig{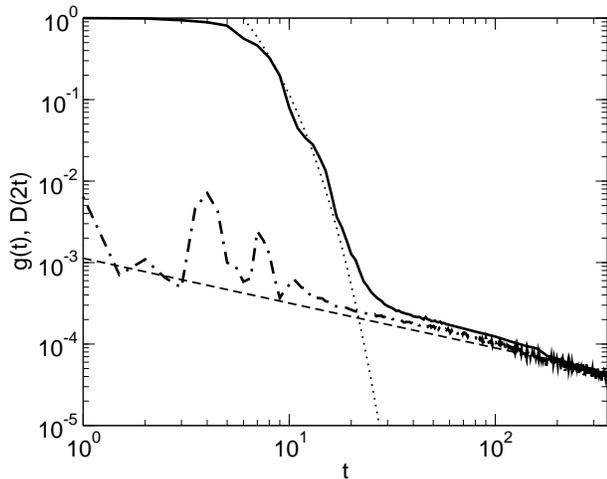}}
\caption{Decay of fidelity for the kicked rotator with
$K_0=2.5$, $L=1$, and $\epsilon=10^{-3}$ (full
curve). The support of the
initial (characteristic) density is inside the chaotic 
component, with $(\theta,p)\in[0,0.2]
\times[0,0.2]$. The dotted curve represents the exponential
decay at a rate given by the Lyapunov exponent 
$\lambda\approx 0.534$. 
The dashed line has slope $-0.55$.
The dot-dashed curve gives the correlation decay $D(2t)$, for 
the same initial density and for twice the time $t$. It is clearly 
seen that, asymptotically, fidelity and correlations have the 
same power law decay $\propto t^{-0.55}$. 
\label{fig:rotfid}}
\end{figure}

Further confirmation for the validity of the above 
illustrated scenario has been obtained by analyzing 
systems in which the asymptotic decay of correlations 
is algebraic. This happens in two cases:
\newline
(i) when the system possesses marginally stable orbits.
A typical example is the stadium billiard in which,
as it is known \cite{stadium}, correlations decay as 
$1/t$; 
\newline
(ii) when there is mixed phase space \cite{mixed}. 
A typical example is the kicked rotator model 
(described by Eq. (\ref{saw}) with $F_0=K_0 \sin \theta$). 
\newline 
Since in the long time limit  
the fidelity decay at time $t$ is still related to the decay 
of correlations at time $2t$, in the case of power law decay 
of correlations as $t^{-\alpha}$, we expect a power law decay 
of fidelity with the same exponent $\alpha$. This is indeed 
confirmed by our numerical results. In Fig. 4 it is shown 
that, for the stadium billiard, fidelity decays asymptotically  
$\propto 1/t$, as expected. In Fig. 5 we compare the fidelity 
decay (at time $t$) and the correlations decay (at time $2t$) 
for the kicked rotator with kicking parameter $K_0=2.5$, 
for which the phase space contains chaotic components and stable 
islands. The correlator is given by  
$D(t)=(C(t)-C(\infty))/
(C(0)-C(\infty))$, with $C(t)=\int_\Omega 
d{\bf x} \rho_0 ({\bf x}, t) \rho ({\bf x},0)$.   
It is seen that, after an initial Lyapunov decay, fidelity 
approaches the same asymptotic power law decay of correlations
\cite{note}.

In summary, we have shown that in chaotic systems 
the asymptotic decay of classical fidelity, 
which describes the structural stability of motion 
under system's perturbations, is analogous to the 
asymptotic decay of correlation functions. 
This asymptotic decay can be either exponential or 
algebraic, depending on the dynamical properties of 
the system. In any instance, it is not related to 
the local exponential instability ruled by the 
Lyapunov exponent and it is $\epsilon$-independent.  
It would be interesting to understand what are 
the implications of these findings for the 
decay of quantum fidelity. 

We thank Roberto Artuso for useful discussions. 
This work was supported in part by the EC RTN contract 
HPRN-CT-2000-0156, the NSA and ARDA under 
ARO contract No. DAAD19-02-1-0086, the
PA  INFM ``Weak chaos: Theory and applications'',
and the PRIN 2000 ``Chaos and localization in classical 
and quantum mechanics''.

\end{multicols}


\begin{thebibliography}{99} 
%\bibitem{a} (a) e-mail:giuliano.benenti@uninsubria.it
%\bibitem{b} (b) e-mail:giulio.casati@uninsubria.it \nonumber
%\bibitem{c} (c) e-mail:gregor.veble@uni-mb.si
\bibitem{peres} A. Peres, Phys. Rev. A {\bf 30}, 
1610 (1984). 
\bibitem{cambridge} G. Casati and B.V. Chirikov, 
{\it Quantum Chaos} (Cambridge University Press, 
Cambdridge, England, 1995). 
\bibitem{arrow} G. Casati, B.V. Chirikov, I. Guarneri,
and D.L. Shepelyansky, Phys. Rev. Lett. {\bf 56},
2437 (1986). 
\bibitem{jalabert} R.A. Jalabert and H.M. Pastawski, 
Phys. Rev. Lett, {\bf 86}, 2490 (2001). 
\bibitem{pastawski} F.M. Cucchietti, H.M. Pastawski, and 
D.A. Wisniacki, Phys. Rev. E {\bf 65}, 045206(R) (2002). 
\bibitem{jacquod} Ph. Jacquod, P.G. Silvestrov, and 
C.W.J. Beenakker, Phys. Rev. E {\bf 64}, 055203(R) (2001), 
\bibitem{tomsovic} N.R. Cerruti and S. Tomsovic, 
Phys. Rev. Lett. {\bf 88}, 054103 (2002). 
\bibitem{prosen} T. Prosen and M. \v Znidari\v c, 
J. Phys. A {\bf 35}, 1455 (2002). 
\bibitem{felix} V.V. Flambaum and F.M. Izrailev, Phys. Rev. E 
{\bf 64}, 036220 (2001). 
\bibitem{zurek} Z.P. Karkuszewski, C. Jarzynski, and 
W.H. Zurek, Phys. Rev. Lett. {\bf 89}, 170405 (2002).  
\bibitem{cohen} D.A. Wisniacki and D. Cohen, 
Phys. Rev. E {\bf 66}, 046209 (2002).
\bibitem{PRE} G. Benenti and G. Casati, Phys. Rev. E 
{\bf 65}, 066205 (2002).
\bibitem{eckhardt} B. Eckhardt, preprint nlin.CD/0210061,
to be published in J. Phys. A.  
\bibitem{ruelle} D. Ruelle, Phys. Rev. Lett. {\bf 56},
405 (1986). 
\bibitem{haake} J. Weber, F. Haake, and P. \v Seba, 
Phys. Rev. Lett. {\bf 85}, 3620 (2000).
\bibitem{fishman} M. Khodas, S. Fishman, and O. Agam, 
Phys. Rev. E {\bf 62}, 4769 (2000). 
\bibitem{stadium} F. Vivaldi, G. Casati, and I. Guarneri, 
Phys. Rev. Lett. {\bf 51}, 727 (1983). 
\bibitem{mixed} C.F.F. Karney, Physica {\bf 8D}, 360 (1983); 
B.V. Chirikov and D.L. Shepelyansky, Physica {\bf 13D}, 
395 (1984). 
\bibitem{note}
We note that the short time Lyapunov decay 
of fidelity is by no means a typical feature of  
correlation functions, since the short time decay of 
$D(t)$ is determined by the motion of the ``center of mass''
of the initial distribution $\rho ({\bf x},0)$, 
a trivial effect which is suppressed in fidelity 
due to the backward evolution. 
\end{thebibliography}
\end{document}